%% file: manuscript.tex
\documentclass[publish,JACIII,paper]{jaciiiarticle}
\usepackage{graphicx}
\usepackage{epsfig}
\usepackage{multirow}
\usepackage{fancybox}
\usepackage{ascmac}
\usepackage{here}
\usepackage[hang,small,bf]{caption}
\usepackage[subrefformat=parens]{subcaption}
\SetVolumeNumber{Vol.0 No.0, 200x} 
\SetArticleID{jc*-**-**-****:}
\begin{document}

\pagestyle{jaciiistyle}

\title{Can Large Language Models Assess Serendipity in Recommender Systems?}
\author{Yu Tokutake and Kazushi Okamoto$^{\dagger}$}
\address{Graduate School of Informatics and Engineering, The University of Electro-Communications\\
		 1-5-1 Chofugaoka, Chofu, Tokyo 182-8585, Japan\\
         E-mail: \{tokutakeyuu, kazushi\}@uec.ac.jp\\
         $^{\dagger}$Corresponding author}
\markboth{Tokutake, Y. and Okamoto, K.}{Can LLMs Assess Serendipity in RSs?}
\dates{00/00/00}{00/00/00}
\maketitle

\begin{abstract}% Do not delete this percent symbol
\input{tex/abstract.tex}
\end{abstract}

\begin{keywords}
recommender system, serendipity, large language model, value judgement
\end{keywords}

\input{tex/section1.tex}
\input{tex/section2.tex}
\input{tex/section3.tex}
\input{tex/section4.tex}
\input{tex/section5.tex}
\input{tex/section6.tex}

\input{tex/acknowledgement.tex}

\input{tex/appendix.tex}

\vspace{25pt}

\input{tex/profiles}
\end{document}

%% file: tex/abstract.tex
\noindent Serendipity-oriented recommender systems aim to counteract over-specialization in user preferences.
However, evaluating a user's serendipitous response towards a recommended item can be challenging because of its emotional nature.
In this study, we address this issue by leveraging the rich knowledge of large language models (LLMs), which can perform a variety of tasks.
First, this study explored the alignment between serendipitous evaluations made by LLMs and those made by humans.
In this investigation, a binary classification task was given to the LLMs to predict whether a user would find the recommended item serendipitously.
The predictive performances of three LLMs on a benchmark dataset in which humans assigned the ground truth of serendipitous items were measured.
The experimental findings reveal that LLM-based assessment methods did not have a very high agreement rate with human assessments.
However, they performed as well as or better than the baseline methods.
Further validation results indicate that the number of user rating histories provided to LLM prompts should be carefully chosen to avoid both insufficient and excessive inputs and that the output of LLMs that show high classification performance is difficult to interpret.

%% file: tex/section1.tex
\section{Introduction} \label{sec:introduction}
Recommender systems (RSs) select items that match the user's preferences from a vast amount of information and help users make effective decisions.
Traditional approaches, such as collaborative filtering, primarily focus on enhancing the recommendation accuracy.
However, these approaches cause over-specialization, in which users are discouraged from exploring items that diverge from their established preferences\cite{2021mar_r.j.Ziarani, 2023aug_z.Fu}.
To tackle this challenge, serendipity-oriented RSs were introduced to suggest items that offer serendipitous experiences to users\cite{2013feb_a.Said, 2015jun_q.Zheng, 2021dec_r.j.Ziarani, 2021oct_m.Zhang, 2023jul_z.Fu, 2018dec_d.Kotkov, 2023oct_y.Tokutake}.
Nonetheless, defining and evaluating serendipity in RSs is challenging because of their subjective and emotional nature\cite{2021dec_r.j.Ziarani}.
Recommendation accuracy is usually assessed through user feedback, which encompasses metrics such as views, clicks, and ratings.
In contrast, serendipity is often unobserved, making a comparison of the performance of serendipity-oriented RSs.
Ideally, we would like to conduct a user survey to obtain feedback on the serendipitous nature of the recommendation results.
Some studies have already implemented such surveys\cite{2023jul_z.Fu, 2018apr_d.Kotkov}, but they are costly, and the responses may not be consistent depending on temporal and psychological factors.
Hence, an assessment method that can serve as a substitute for human assessment is required.

In this context, large language models (LLMs) have emerged as a promising solution.
With their remarkable capabilities in various tasks, including natural language processing (NLP), LLMs can offer a versatile framework for assessing serendipity in different RSs, potentially reducing reliance on human surveys.
In the realm of RS research, LLM4Rec, which leverages LLMs' extensive knowledge and inference capabilities of LLMs for RSs, has been proposed.
Hua et al. \cite{2023sep_w.Hua} argued that LLM4Rec can be categorized into three types: (1) LLMs as RSs\cite{2023apr_j.Liu}; (2) LLMs in RSs, such as feature extractors\cite{2023jun_y.Xi}; and (3) RSs in LLMs, such as recommendation agents\cite{2023oct_a.Zhang}.
In this study, we focused on the latter two types: (2) and (3).
Recent studies have investigated the utilization of LLMs to enhance beyond-accuracy metrics.
For instance, Carraro et al.\cite{2024jan_d.Carraro} introduced a re-ranking technique employing LLMs to improve the diversity of the recommendation results.
Their findings indicated that LLMs were able to interpret the re-ranking task, but had a lower trade-off between relevance and diversity than the baseline method.
As far as we know, the application of LLMs in serendipitous assessment remains unexplored.

In this study, we investigated the potential of LLMs in assessing serendipity in RSs, focusing on the following research questions:
\begin{enumerate}
  \item[\textbf{RQ1:}] To what extent do LLM-based serendipitous assessments align with human assessment?
  \item[\textbf{RQ2:}] Can LLMs assessments enhance performance relative to baseline methods?
\end{enumerate}
To address these inquiries, we propose a serendipitous assessment method using a LLM and conduct an experiment using a benchmark dataset with human-provided ground truths for serendipity.
The LLM-based method determines whether a recommended item is serendipitous, based on the users' rating history.
In the experiment, we evaluate the classification performance of the proposed and baseline methods by applying different LLMs and their prompts to the proposed method.
The information provided in the prompts varied, including item names only, pairs of item names and ratings, pairs of item names and genres, and pairs of item names, ratings, and genres.

The remainder of this paper is organized as follows.
Section \ref{sec:related_work} reviews the studies on serendipity in recommender systems and LLMs for assessment.
Section \ref{sec:proposed_method} introduces the LLM-based assessment method.
Section \ref{sec:experiment} describes an experiment using a benchmark dataset.
Section \ref{sec:result} analyzes and discusses the experimental results in terms of the classification performance, rating history size used, and interpretation of LLMs outputs.

%% file: tex/section2.tex
\section{Related Work} \label{sec:related_work}
\subsection{Serendipity in Recommender Systems} \label{subsec:ser_rs}
In RS studies, the concept of the serendipity has garnered significant interest over time, with various RSs designed to enhance the serendipity across multiple domains\cite{2021mar_r.j.Ziarani}.
While there is no unified definition of serendipity, previous studies have identified common components, such as relevance, unexpectedness, novelty, and diversity\cite{2023mar_d.Kotkov, 2023aug_z.Fu}.
Serendipity-oriented RSs integrate these components into their algorithms, which include $k$-nearest neighbor\cite{2013feb_a.Said}, matrix factorization\cite{2015jun_q.Zheng}, convolutional neural networks\cite{2021dec_r.j.Ziarani}, Transformer\cite{2021oct_m.Zhang, 2023jul_z.Fu}, and re-ranking for recommendation lists generated by accuracy-oriented algorithms\cite{2018dec_d.Kotkov, 2023oct_y.Tokutake}.

Evaluation methods for assessing serendipity also vary among studies, with several metrics have been proposed.
For instance, Adamopoulos et al.\cite{2014dec_p.Adamopoulos} defined serendipity in a recommendation list as the proportion of useful yet unexpected items among recommended items.
Yu et al.\cite{2017jul_h.Yu} determined the serendipity of a recommended item by calculating the ratio of other users’ ratings for the item to the similarity-weighted ratings given to other items by the user.
Some studies have assessed serendipity based on interaction scores between users and items, as defined in RS algorithms\cite{2021oct_m.Zhang, 2023oct_y.Tokutake}.
In response to these current definitions and evaluations of serendipity, Kotkov et al.\cite{2023mar_d.Kotkov} broadened the definition within the context of RSs to reconcile the discrepancies between specific RS definitions and a generalized understanding of serendipity.
Additionally, they devised an experimental design that evaluated both initial and achieved goals pre- and post-recommendation, countering the previous focus solely on the latter period.

The challenge of validating serendipity-oriented RSs is compounded by the lack of ground-truth data on whether users perceive the recommended item as serendipitous in many benchmark datasets.
However, some studies gathered such data through user surveys\cite{2023jul_z.Fu, 2018apr_d.Kotkov}.
Presently, the Serendipity-2018 dataset\footnote{https://files.grouplens.org/datasets/serendipity-sac2018/} stands as the sole publicly available dataset that collects feedback from MovieLens users regarding recommended movies, as explored by Kotkov et al.\cite{2018apr_d.Kotkov}.

\subsection{Large Language Models for Assessment} \label{subsec:llm_assess}
Recent studies in NLP have turned to LLMs to evaluate natural language generation (NLG) tasks instead of traditional automatic evaluation metrics \cite{2023dec_j.Wang, 2023jun_l.Zheng}.
For instance, \cite{2023jun_l.Zheng} employed a benchmark dataset of multi-category questions and found that the agreement between the outputs of LLMs, as evaluated by a human expert, and those compared using GPT-4\footnote{https://platform.openai.com/docs/models/gpt-4-and-gpt-4-turbo} exceeded 80\%.
This level of agreement was comparable to that observed when two human evaluators were compared.

In addition to NLG tasks, LLMs have found applications in assessing information retrieval\cite{2023aug_g.Faggioli, 2023dec_w.Sun} and legal judgement prediction\cite{2022dec_d.Trautmann, 2023jun_c.Jiang}.
Faggioli et al.\cite{2023aug_g.Faggioli} employed LLMs to evaluate the relevance between a given topic and a document, and found that relevance judgments made by LLMs were highly correlated with human judgments.
Trautmann et al.\cite{2022dec_d.Trautmann} employed LLMs for a binary classification task to determine whether a case violates human rights provisions.
The experimental results demonstrated that while their proposed method outperformed the baseline methods using majority voting and random selection but was inferior to the method that employs supervised learning.
Jiang et al.\cite{2023jun_c.Jiang} enhanced performance by introducing a legal syllogism into the LLMs' prompt, and outputting the inference process, including the legal text and the facts of the case, and judgements.

%% file: tex/section3.tex
\section{LLM-based Serendipitous Assessment} \label{sec:proposed_method}
We propose a method for serendipitous assessment using LLMs, incorporating the user's rating history and recommended items.
The framework of our proposed method is illustrated in \textbf{Fig. \ref{figure:framework}}.

\input{figure/framework.tex}

\subsection{Task Definition}
The objective of this study provides user $u$'s rating history $I_{u}$ and query item $i$, and determines whether item $i$ is serendipitous for user $u$ or not.
Thus, the object function is $f$, which is defined as follows:
\begin{equation}
    f(I_{u}, i) = 
    \begin{cases}
        1 & \text{if $i$ is serendipitous for $u$} \\
        0 & \text{otherwise}.
    \end{cases} \label{eq:obf_binary}
\end{equation}
In addition, there is a case in which the objective function $f$ outputs the degree of serendipity, and $f$ is defined as follows:
\begin{equation}
    f(I_{u}, i) = s, \quad s \in [0, 1] \label{eq:obf_cont}
\end{equation}
in this case, and alternatively.
In Eq. (\ref{eq:obf_cont}), the threshold value was employed to convert it into a binary representation, as shown in Eq. (\ref{eq:obf_binary}).
Typically, RSs present multiple recommended items in list format.
In contrast, we simplified the problem by assessing the serendipity of a single item.
Consequently, when assessing the serendipity of the entire recommendation list, function $f$ is applied individually to each item.

In this study, we introduce $f_{\mathrm{LLM}}$ with LLMs for $f$.
The function $f_{\mathrm{LLM}}$ generates a prompt based on the information from the rating history $I_u$ and query item $i$ and receives the output of the LLMs.
This output is a binary value, $f_{\mathrm{LLM}}$, corresponding to Eq. (\ref{eq:obf_binary}).

\subsection{Prompt for LLMs}
The prompt utilized for the LLMs is depicted in \textbf{Fig. \ref{figure:prompt}}.
Employing a few-shot learning approach, one positive and one negative example were provided to the LLMs.
As discussed in Section \ref{sec:related_work}, there are various definitions of serendipity in RSs.
However, in this study, we do not provide these definitions and focus solely on the term ``serendipity'' in order to explore LLMs' knowledge and inference capabilities of the LLMs.

The prompt can include different types of information.
We propose four variations based on the item name, user rating of the item, and item genre.
\begin{itemize}
\item{implicit: only item names}
\item{explicit: item names and users' ratings}
\item{implicit with genres: item names and genres}
\item{explicit with genres: item names, ratings, and genres}
\end{itemize}
It is important to note that in explicit cases, the rating of the recommended item is considered a predicted value by RSs, rather than an actual user rating observed.
Each item is represented in a tuple and list format, arranged from oldest to newest (i.e., \texttt{(item\_name, rating, [genre1, genre2, ...]), ...}).
Specifically, `user\_rated\_movies\_example' is ``War Dogs, Gosford Park, ...'' for implicit, ``(War Dogs, 3.5), (Gosford Park, 3.0), ...'' for explicit, ``(War Dogs, Comedy), (Gosford Park, [Comedy, Drama, Mystery]), ...'' for implicit with genres, and ``(War Dogs, 3.5, Comedy), (Gosford Park, 3.0, [Comedy, Drama, Mystery]), ...'' for explicit with genres.

\input{figure/prompt.tex}

%% file: figure/framework.tex
\begin{figure}
    \centering
    \epsfig{file=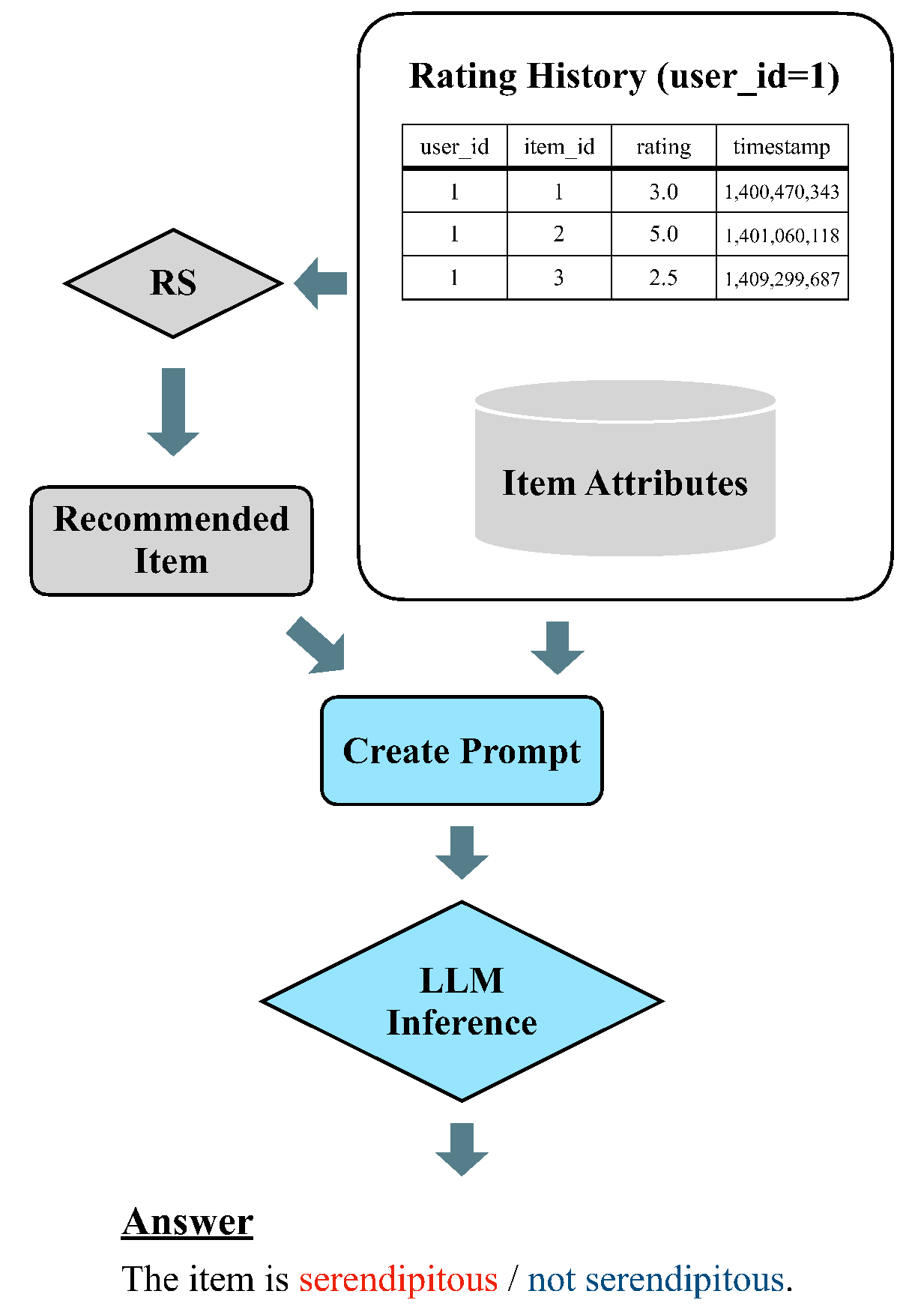, width=82mm}
    \caption{The framework of the proposed method}
    \label{figure:framework}
\end{figure}

%% file: figure/prompt.tex
\begin{figure}[t]
\centering
\footnotesize{
\begin{screen}
    Please judge whether `recommended\_movie' is serendipitous or not given `user\_rated\_movies'. \\\\
   
    \#\# Background \\
    * You use a movie rating platform and have rated some movies. \\
    * Now the movie is recommended based on your rating history from the platform. \\
    * You are given the \{title, rating, genres\} of the recommended movie and rated movies. \\
    * The rating history is comma-separated and sorted from oldest to newest. \\\\
    
    \#\# Output Format \\
    * You should answer just `Yes' or `No' after the `is\_serendipitous: ' prefix.\\
    * Generate only the requested output, don't include any other language before or after the requested output. \\\\
    
    \#\# Examples \\
    \#\#\# Example 1 \\
    user\_rated\_movies: \{user\_rated\_movies\_example1\} \\
    recommended\_movie: \{recommended\_movie\_example1\} \\
    is\_serendipitous: Yes \\\\
    
    \#\#\# Example 2 \\
    user\_rated\_movies: \{user\_rated\_movies\_example2\} \\
    recommended\_movie: \{recommended\_movie\_example2\} \\
    is\_serendipitous: No \\\\
    
    \#\# Response \\
    user\_rated\_movies: \{user\_rated\_movies\} \\
    recommended\_movie: \{recommended\_movie\} \\
    is\_serendipitous:
\end{screen}
}
\caption{Prompt for LLMs}
\label{figure:prompt}
\end{figure}

%% file: tex/section4.tex
\section{Experiment} \label{sec:experiment}
An experiment was conducted using a benchmark dataset to evaluate the performance of the proposed method.

\subsection{Dataset and Preprocessing}
The Serendipity-2018 dataset mentioned in Section \ref{subsec:ser_rs} was used in this experiment.
The statistics for the dataset are presented in \textbf{Table \ref{table:statistics}}.
User ratings were assigned in increments of 0.5 stars, ranging from 0.5 to 5.
Movies were categorized into 18 genres, such as Action or Sci-Fi, with the option of assigning multiple genres to a single movie.
In addition to the users, items, and ratings displayed in \textbf{Table \ref{table:statistics}}, this dataset included a total of 481 users who responded to the questionnaire outlined in \textbf{Table \ref{table:questions}}, with a maximum of five item responses per individual.
Each question was rated on a scale of 1 to 5 stars (1 = strongly disagree, 2 = disagree, 3 = neither agree nor disagree, 4 = agree, and 5 = strongly agree).

Following methodologies from previous studies \cite{2021dec_r.j.Ziarani, 2018dec_d.Kotkov, 2018apr_d.Kotkov}, we generated a binary label to indicate whether a user perceived an item as serendipitous. 
As outlined by Kotov et al.\cite{2018apr_d.Kotkov}, we assigned a binary ground-truth label to each response using the rule provided in the appendix.
This process yielded 277 serendipitous and 1,873 non-serendipitous feedback instances.

\input{table/statistics.tex}
\input{table/questions.tex}

\subsection{LLMs for Verification}
For validation, we employed three models: GPT-3.5\footnote{https://platform.openai.com/docs/models/gpt-3-5} (gpt-3.5-turbo-0613 version) and GPT-4 (gpt-4-0613 version) from OpenAI and Llama2-13B-Chat\footnote{https://huggingface.co/meta-llama/Llama-2-13b-chat-hf} from Meta.
The temperature was set to 0.0 for all three models.
The upper limit of the LLM context length was restricted to the inclusion of a user's entire rating history, and we used only the 10 most recent items as input for the prompts.

\subsection{Baseline Methods}
We employed six methods, categorized into two types: traditional and serendipity-oriented RS methods, as baselines for comparison with the proposed method.
In this experiment, the supervised method was not utilized as the baseline method due to insufficient training data and a lack of ground truth for serendipitous items when applied to actual RSs.

\subsubsection{Traditional Methods}
Traditional methods consist of three approaches for label prediction: outputting all negatives (\textbf{all neg.}) $f_{all\_neg}$, outputting all positives (\textbf{all pos.}), $f_{all\_pos}$ and randomly outputting the labels (\textbf{random}) $f_{random}$.
Functions $f_{all\_neg}$ and $f_{all\_pos}$ are defined as 
\begin{equation}
    f_{all\_neg}(I_{u}, i) = 0, \quad f_{all\_pos}(I_{u}, i) = 1.
\end{equation}

\subsubsection{Serendipity-oriented RS Methods}
We employed the Serendipity-Oriented Greedy (SOG) algorithm \cite{2018dec_d.Kotkov}, a re-ranking algorithm that sorts the recommendation list obtained using Singular Value Decomposition (SVD).
The re-ranking score, $score_{uiB}$, is calculated as follows:
\begin{align}
  score_{uiB} =\ &\alpha_{rel}\cdot \hat{r}_{ui} + \alpha_{div}\cdot {div}_{iB} + \alpha_{prof}\cdot {prof}_{ui} \nonumber \\ &+ \alpha_{unpop}\cdot {unpop}_{i} \label{eq:sog_score}
\end{align}
where $\hat{r}_{ui}$ is the predicted rating for item $i$ by user $u$ using SVD, ${div}_{iB}$ is the average content dissimilarity between items in candidate list $B$ and item $i$, ${prof}_{ui}$ is the average content dissimilarity between items rated by user $u$ in the past and item $i$, and ${unpop}_{i}$ is the unpopularity of item $i$ calculated as the percentage of users who did not rate it.
In this case, because ${div}_{iB}$ cannot be defined owing to the absence of a candidate list $B$, $score_{uiB}$ is calculated with ${div}_{iB} = 0$.
Therefore, we used three methods: $\textbf{score}$, $\textbf{prof}$, and $\textbf{unpop}$.
Following Kotkov et al.\cite{2018dec_d.Kotkov}, we set $\alpha_{rel}=0.9$, $\alpha_{prof}=0.7$ and $\alpha_{unpop}=0.7$ for $\textbf{score}$.
When used as discriminators in this study, they are calculated as follows:
\begin{equation}
    f_{score}(I_{u}, i) = 0.9 \cdot \hat{r}_{ui} + 0.7 \cdot {prof}_{ui} + 0.7 \cdot {unpop}_{i}
\end{equation}
\begin{equation}
    f_{prof}(I_{u}, i) = {prof}_{ui}
\end{equation}
\begin{equation}
    f_{unpop}(I_{u}, i) = {unpop}_{i}.
\end{equation}
Using both the proposed LLM-based method and the traditional baseline methods, the outputs were binary, indicating either a positive or negative result.
For $f_{score}$, $f_{prof}$, and $f_{unpop}$, whose outputs are continuous values, we transformed the outputs into binary values, as in the other methods, by establishing thresholds.
The thresholds are the values at which the top $q$\% ($5\leq q \leq 95$ in increments of 1) are positive.
In simpler terms, these thresholds represent the values in the $q$ th quartile.

\subsection{Evaluation Metrics}
This experiment aimed to evaluate the classification performance of each method in a binary classification task.
Therefore, we utilized accuracy, precision, recall, and F1-score as evaluation metrics.
It is crucial to note that precision, recall, and F1-score are macro metrics that incorporate negative cases in their calculations to mitigate distributional bias of the ground truth.
The ground truth indicating whether user $u$ perceives serendipity for item $i$ is denoted by $g\ (\in\{0, 1\})$.
Each metric was calculated as follows.
\begin{equation}
    \mathrm{Accuracy}: \frac{TP + TN}{TP + TN + FP + FN}
\end{equation}
\begin{equation}
    \mathrm{Precision}: \frac{1}{2} (\frac{TP}{TP + FP} + \frac{TN}{TN + FN})
\end{equation}
\begin{equation}
    \mathrm{Recall}: \frac{1}{2} (\frac{TP}{TP + FN} + \frac{TN}{TN + FP})
\end{equation}
\begin{equation}
    \mathrm{F1\mathchar`-score}: \frac{TP}{2TP + FP + FN} + \frac{TN}{2TN + FP + FN}
\end{equation}
where $TP$ represents the number of items with $f = 1$ and $g = 1$; $FP$ represents the number with $f = 1$ and $g = 0$; $TN$ is the number with $f = 0$ and $g = 0$; and $FN$ represents the number with $f = 0$ and $g = 1$.

%% file: table/statistics.tex
\begin{table}[t]
  \centering
  \caption{Statistics of Serendipity-2018}
  \label{table:statistics}
  \begin{tabular}{ccccc}
    \hline
    users & items & ratings & genres & feedbacks \\
    \hline
    \hline
    9,997,850 & 104,661 & 49,151 & 18 & 2,150 \\
    \hline
  \end{tabular}
\end{table}

%% file: table/questions.tex
\begin{table*}[t]
  \centering
  \caption{Question description of Serendipity-2018}
  \label{table:questions}
  \begin{tabular}{cl}
    \hline
    No. & \multicolumn{1}{c}{Description} \\
    \hline
    \hline
    Q1 & The first time I heard of this movie was when MovieLens suggested it to me. \\
    \hline
    Q2 & MovieLens influenced my decision to watch this movie. \\
    \hline
    Q3 & I expected to enjoy this movie before watching it for the first time. \\
    \hline
    \multirow{2}{*}{Q4} & This is the type of movie I would not normally discover on my own; \\
    & I need a recommender system like MovieLens to find movies like this one. \\
    \hline
    Q5 & This movie is different (e.g., in style, genre, topic) from the movies I usually watch. \\
    \hline
    Q6 & I was (or, would have been) surprised that MovieLens picked this movie to recommend to me. \\
    \hline
    Q7 & I am glad I watched this movie. \\
    \hline
    Q8 & Watching this movie broadened my preferences. Now I am interested in a wider selection of movies. \\
    \hline
  \end{tabular}
\end{table*}

%% file: tex/section5.tex
\section{Result and Discussion} \label{sec:result}
In this section, we analyze the experimental results for RQ1 and RQ2.
Additionally, we confirmed the change in classification performance depending on the size of the rating history used and attempted to interpret the LLMs outputs.
The results of the classification performance are summarized are shown in \textbf{Table \ref{table:accuracy_result}}.
For the SOG score values (i.e., $prof$, $unpop$, and $score$), the maximum value for each metric is reported as the threshold that varies, representing the theoretical value.
The best value for each metric is highlighted in bold, and the second-best value is underlined.

\input{table/accuracy_comparison.tex}

\subsection{Analysis for RQ1}
From \textbf{Table \ref{table:accuracy_result}}, the combinations of LLMs and prompts for the proposed method, namely GPT-3.5 (explicit w/ genres), Llama2-13B-Chat (implicit), GPT-3.5 (implicit w/ genres), and Llama2-13B-Chat (implicit w/ genres), achieved the second highest score for accuracy, the highest scores for recall, and the highest and second highest scores for F1-score, respectively.
Accuracy reached over 0.8 in GPT-3.5 (explicit w/ genres), but most other metrics ranged between 0.5 and 0.6.
This indicates a relatively low classification performance based on human assessment.
The experimental setting, including the imbalance in the ground-truth labels, partly contributed to this result.
However, this finding implies that the LLM-based assessments do not have a high rate of agreement with human assessments, indicating that there is room for improvement.

Next, we examined the differences between the methods that performed well and those that did not within the LLM-based assessments.
Since the results in \textbf{Table \ref{table:accuracy_result}} indicate that GPT-4 did not perform well in this experiment, we do not discuss it further.
The confusion matrices for the LLM-based assessment results are depicted in \textbf{Fig. \ref{figure:llama2_confusion_matrix}} and \textbf{Fig. \ref{figure:gpt-3.5_confusion_matrix}}.
Firstly, \textbf{Fig. \ref{figure:llama2_confusion_matrix}} and \textbf{Fig. \ref{figure:gpt-3.5_confusion_matrix}} illustrate that the LLM assessments tended to exhibit bias towards either the positive ($TP + FP$) or negative ($TN + FN$) side.
We compared the confusion matrices of GPT-3.5 (implicit w/ genres) and Llama2-13B-Chat (implicit w/ genres), which exhibited superior F1 scores.
The number of positive assessments ($TP + FP$) tended to be higher than that for the two methods, which was particularly noticeable for Llama2-13B-Chat (implicit, explicit, and implicit w/ genres).
Additionally, $TN$ decreased and $FP$ increased when comparing implicit and explicit, and implicit w/ genres and explicit w/ genres were compared, except for GPT-3.5 (implicit w/ genres) and GPT-3.5 (explicit w/ genres).
These results suggest that utilizing user ratings may lead to an overestimation of positive outcomes.
This may be because of the high predicted rating of the recommended items by SVD, with an average predicted rating of 3.98.
Furthermore, 79.3\% of the items had ratings higher than the user's average rating for the last 10 items, indicating that the LLMs may be overly positive in their predictions.
This inclination is expected to be more pronounced for Llama2-13B-Chat than for GPT-3.5, especially because performance declines for explicit and explicit w/ genres.
Finally, comparing implicit to implicit w/ genres and explicit to explicit w/ genres, either $TP$ or $TN$ increase.
Here, \textbf{Table \ref{table:accuracy_result}} demonstrates that providing item genres improved the F1-score for the LLM-based method, exception for Llama2-13B-Chat (implicit w/ genres) and Llama2-13B-Chat (explicit w/ genres).
These results suggest that explicitly including genre information is beneficial in the serendipitous assessments of LLMs.
These findings indicate that the information provided to the prompt, such as user ratings and item genres, significantly influenced the LLMs' assessments.

\input{figure/llama2_confusion_matrix.tex}
\input{figure/gpt-3.5_confusion_matrix.tex}

\subsection{Analysis for RQ2}
From \textbf{Table \ref{table:accuracy_result}}, Llama2-13B-Chat and GPT-3.5 demonstrate equivalent or superior performance compared to the baseline methods.
Determining an appropriate threshold for SOG scores in an operational RS environment is challenging.
Therefore, we consider the utilization of LLMs for serendipitous assessment significant because their performance is equal to or better than the theoretical value of the SOG scores.
Additionally, LLMs enable a broader range of individuals to assess the recommendation results. 
Users only need to generate a prompt that does not necessitate the evaluation of others, as is the case with $unpop$.
Among the three SOG scores, $prof$, which measures context dissimilarity, yielded the best F1-score, showing a similar trend to LLMs, where the performance is improved by providing genre information.

\subsection{Investigate the Impact of the User Rating History Length Used}
We investigated the change in classification performance when varying $n$, the number of the user's last items given to the LLMs prompts, which was set to a fixed value ($n=10$) in the previous experiment.
Here we focus solely on GPT-3.5 (implicit w/ genres), which attained the highest F1-score in \textbf{Table \ref{table:accuracy_result}}.
The values of each metric when $n$ is varied by $\{3, 5, 10, 20, 50\}$ are illustrated in \textbf{Table \ref{table:accuracy_length_history}}.
As a result, the best values were obtained for all metrics at $n=10$ except for recall, and the classification performance improved as $n$ approached 10 for both $n < 10$ and $n > 10$.
This suggests the necessity to choose the optimal number of most recent user preferences to be referenced in LLMs, neither too short nor too long.

\input{table/accuracy_length_history.tex}

\subsection{Interpret the Outputs of LLMs}
\input{table/pr-auc_logistic_regression.tex}
\input{table/logistic_regression_result.tex}

We interpreted the outputs of the LLMs (black boxes) using the baseline methods.
Specifically, a logistic regression analysis with SOG scores as explanatory variables was performed as follows:
\begin{equation}
     \hat{g}(I_{u}, i) = \frac{1}{1 + \exp \{ -L_{ui}(\boldsymbol{\beta}) \}}
\end{equation}
\begin{equation}
     L_{ui}(\boldsymbol{\beta}) = \beta_{0} + \beta_{1} \cdot \hat{r}_{ui} + \beta_{2} \cdot {prof}_{ui} + \beta_{3} \cdot {unpop}_{i}
\end{equation}
where $\beta_{\ast}$ are the regression coefficients.
Note that all input variables are standardized.
The fitness of the regression model was evaluated using the area under the precision-recall curve (PR-AUC), which was computed from the $\hat{g}$ and the LLM outputs.

First, we discuss the PR-AUC.
\textbf{Table \ref{table:pr-auc}} lists the PR-AUC for each method.
Focusing on Llama2-13B-Chat (implicit w/ genres) and GPT-3.5 (implicit w/ genres and explicit w/ genres), which have PR-AUC in the bottom three, these methods have F1-score in the top three (see \textbf{Table \ref{table:accuracy_result}}).
This indicates that interpreting LLMs outputs with high classification performance using SOG scores is challenging.
This result indicates that LLMs are influenced by factors beyond human explanation; therefore, they should be utilized.

Next, we discuss a regression model for LLMs that can be interpreted using the SOG scores.
\textbf{Table \ref{table:log_reg}} presents the regression results for Llama2-13B-Chat (explicit and explicit w/ genres) and GPT-4 (implicit), where PR-AUC is greater than 0.9, including coefficients (Coef), $p$-values, and 95\% confidence intervals (95\%CI) for each variable.
In \textbf{Table \ref{table:log_reg}}, $p$-values less than 0.05 are underlined.
The result indicates that $prof$ is a significant variable for serendipitous assessments, with a $p$-value below 0.05 and a 95\%CI that does not cross 0 for all three methods.
Because the coefficients of $prof$ are all positive, it suggests that the assessments of the LLMs are expected to be more positive as the context dissimilarity between the rating history and the recommended item increases.
The coefficient of $prof$ in Llama2-13B-Chat (explicit w/ genres) exhibited the largest absolute value among all the explanatory variables and methods.
These statements reinforce the fact that LLMs focus on the context of the items and that the genre information of the items is useful for LLMs' serendipitous assessments of LLMs as discussed earlier.

%% file: table/accuracy_comparison.tex
\begin{table*}[t]
  \centering
  \caption{Classification performance for each method}
  \label{table:accuracy_result}
  \begin{tabular}{lcccc}
    \hline
    \multicolumn{1}{c}{Method} & Accuracy & Precision & Recall & F1-score \\
    \hline\hline
    all neg. & \textbf{0.871} & 0.436 & 0.500 & 0.466 \\
    all pos. & 0.129 & 0.064 & 0.500 & 0.114 \\
    random & 0.490 & 0.497 & 0.494 & 0.413 \\
    \hline
    SOG ($prof$) & 0.834 & \textbf{0.550} & 0.536 & 0.510 \\
    SOG ($unpop$) & 0.826 & 0.521 & 0.517 & 0.479 \\
    SOG ($score$) & 0.831 & \underline{0.545} & \underline{0.548} & 0.491 \\
    \hline
    Llama2-13B-Chat (implicit) & 0.579 & 0.526 & \textbf{0.557} & 0.476 \\
    Llama2-13B-Chat (explicit) & 0.196 & 0.496 & 0.497 & 0.195 \\
    Llama2-13B-Chat (implicit w/ genres) & 0.808 & 0.526 & 0.521 & \underline{0.522} \\
    Llama2-13B-Chat (explicit w/ genres) & 0.151 & 0.484 & 0.496 & 0.143 \\
    \hline
    GPT-3.5 (implicit) & 0.789 & 0.503 & 0.502 & 0.502 \\
    GPT-3.5 (explicit) & 0.635 & 0.509 & 0.517 & 0.483 \\
    GPT-3.5 (implicit w/ genres) & 0.753 & 0.524 & 0.531 & \textbf{0.525} \\
    GPT-3.5 (explicit w/ genres) & \underline{0.840} & 0.538 & 0.516 & 0.513 \\
    \hline
    GPT-4 (implicit) & 0.184 & 0.515 & 0.508 & 0.181 \\
    GPT-4 (explicit) & 0.254 & 0.525 & 0.529 & 0.254 \\
    GPT-4 (implicit w/ genres) & 0.427 & 0.505 & 0.511 & 0.383 \\
    GPT-4 (explicit w/ genres) & 0.455 & 0.500 & 0.499 & 0.397 \\
    \hline
  \end{tabular}
\end{table*}

%% file: figure/llama2_confusion_matrix.tex
\begin{figure}[t]
    \begin{tabular}{cc}
      \begin{minipage}[t]{0.45\hsize}
        \centering
        \epsfig{file=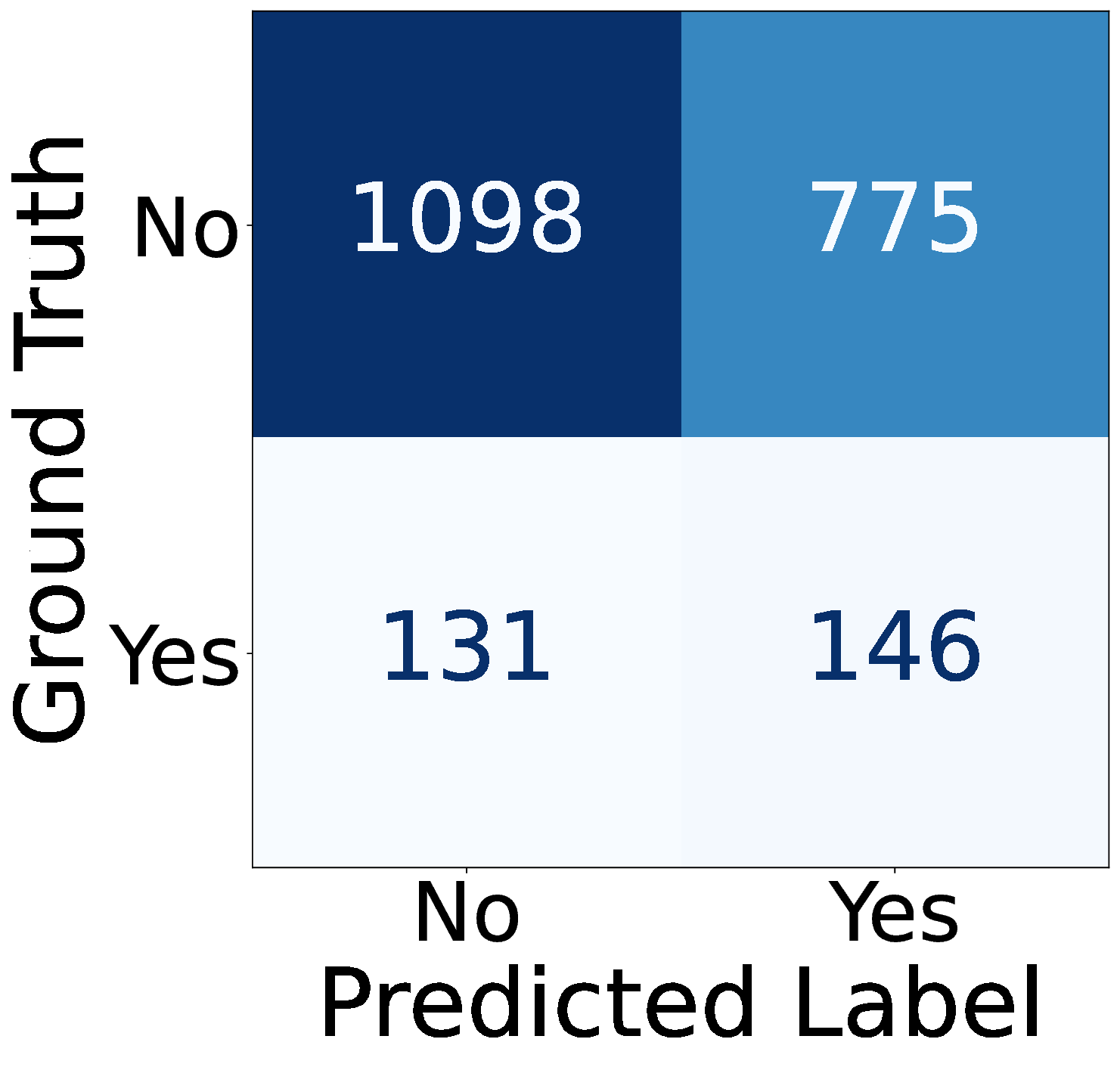, width=34mm}
        \subcaption{implicit}
        \label{subfigure:llama2_implicit_confusion_matrix}
      \end{minipage} &
      \begin{minipage}[t]{0.45\hsize}
        \centering
        \epsfig{file=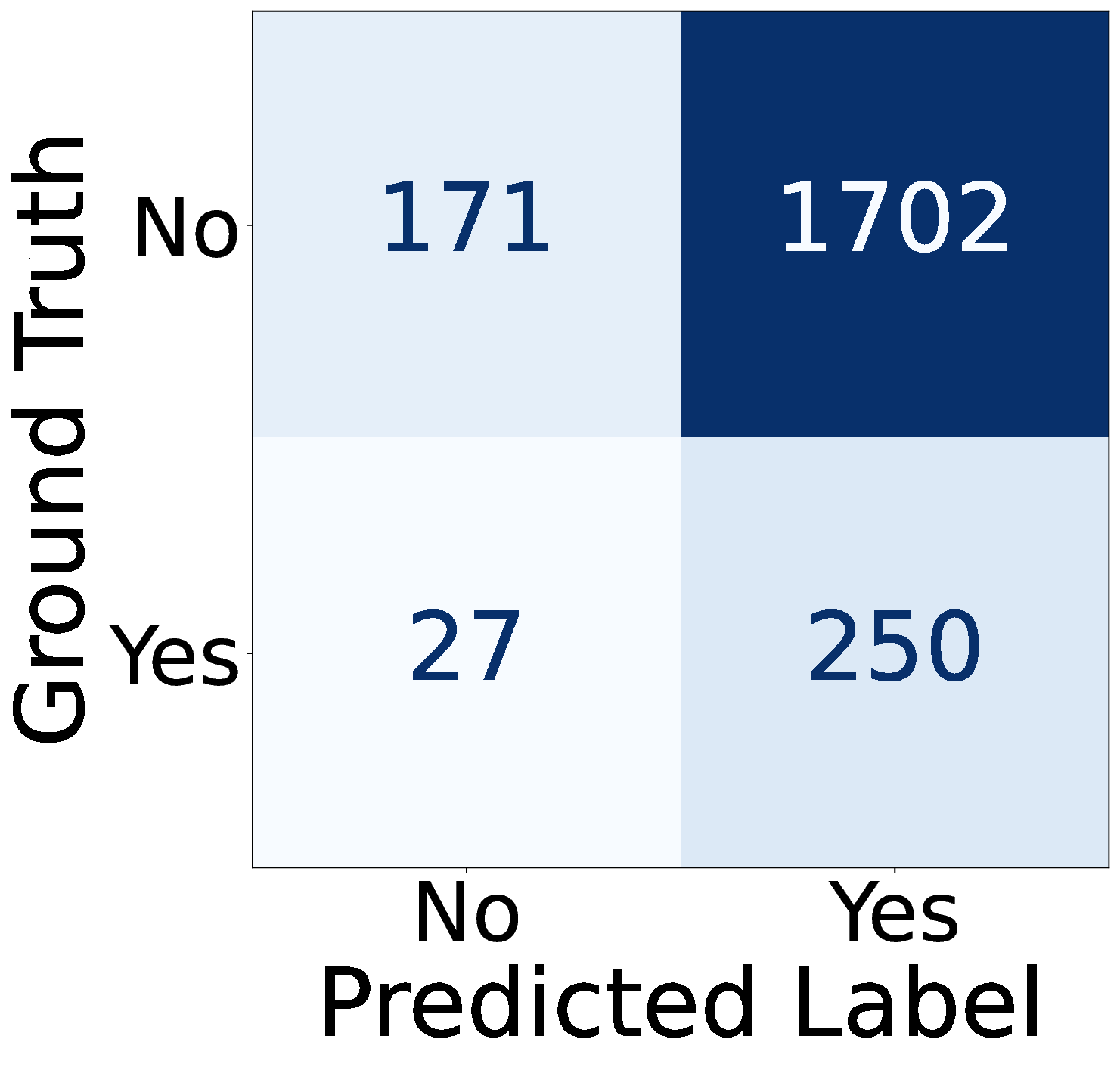, width=34mm}
        \subcaption{explicit}
        \label{subfigure:llama2_explicit_confusion_matrix}
      \end{minipage} \\

      \begin{minipage}[t]{0.45\hsize}
        \centering
        \epsfig{file=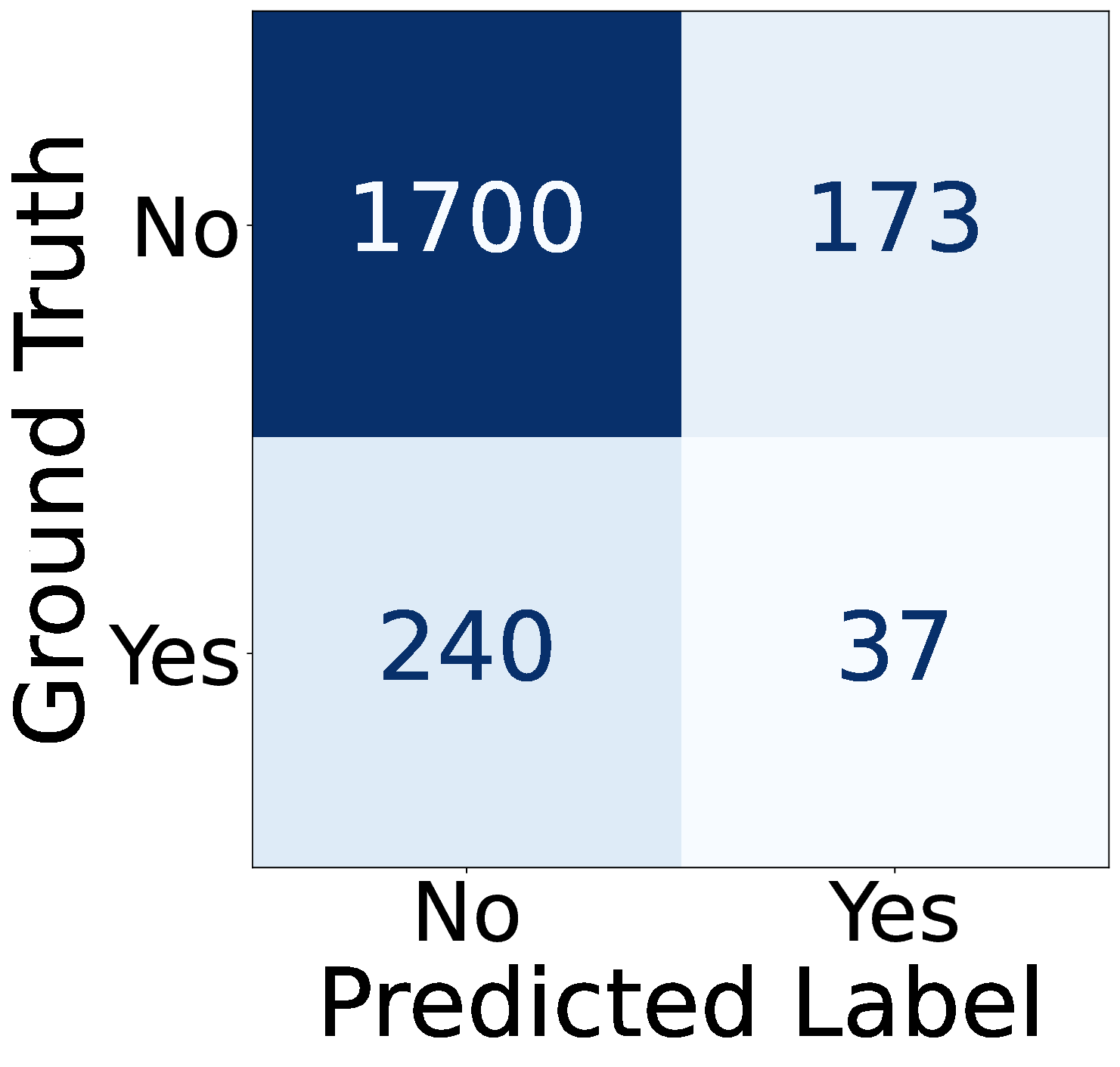, width=34mm}
        \subcaption{implicit w/ genres}
        \label{subfigure:llama2_implicit_with_genres_confusion_matrix}
      \end{minipage} &
      \begin{minipage}[t]{0.45\hsize}
        \centering
        \epsfig{file=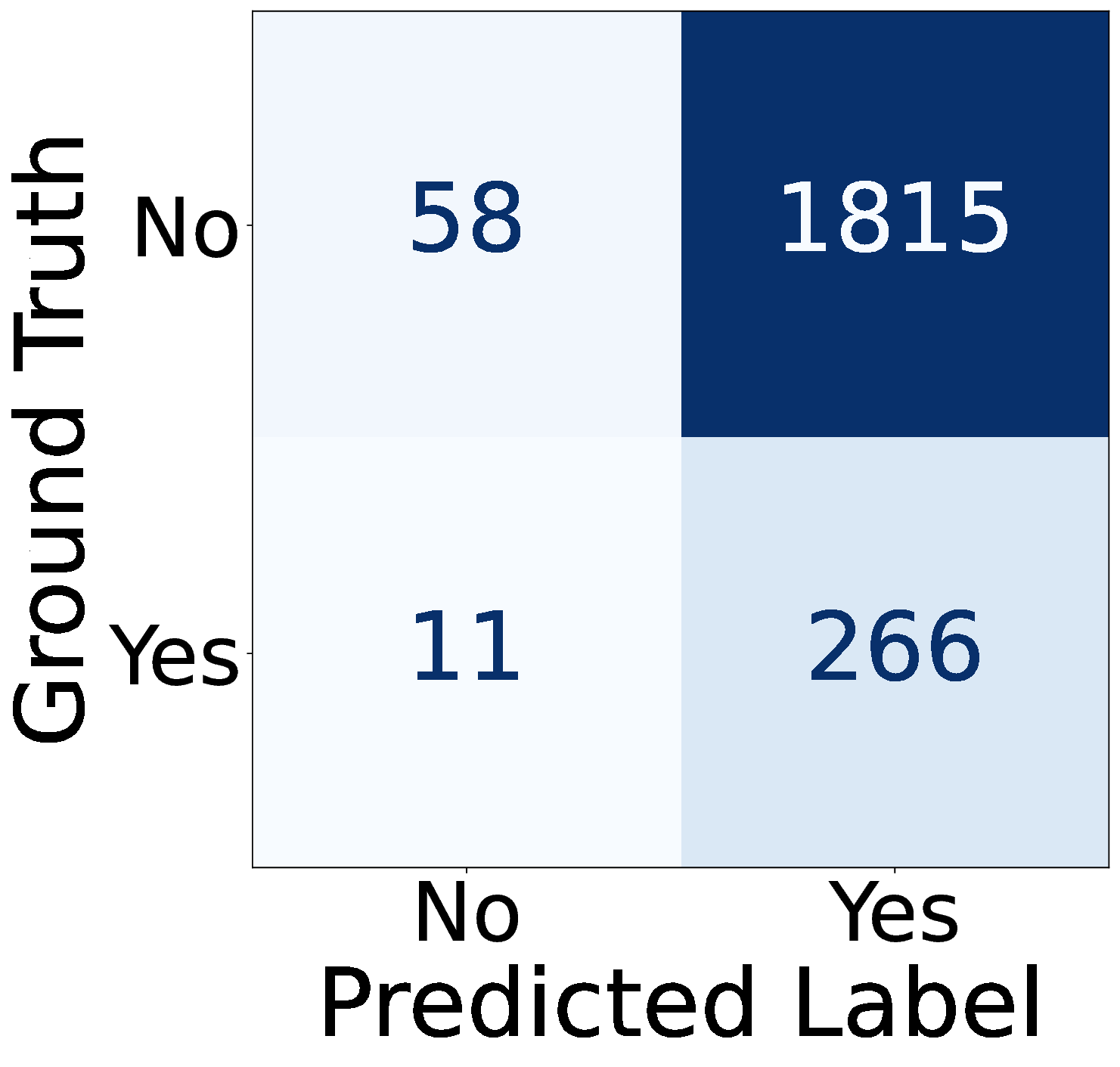, width=34mm}
        \subcaption{explicit w/ genres}
        \label{subfigure:llama2_explicit_with_genres_confusion_matrix}
      \end{minipage}
    \end{tabular}
    \caption{Confusion Matrix - Llama2-13B-Chat}
    \label{figure:llama2_confusion_matrix}
\end{figure}

%% file: figure/gpt-3.5_confusion_matrix.tex
\begin{figure}[t]
    \begin{tabular}{cc}
      \begin{minipage}[t]{0.45\hsize}
        \centering
        \epsfig{file=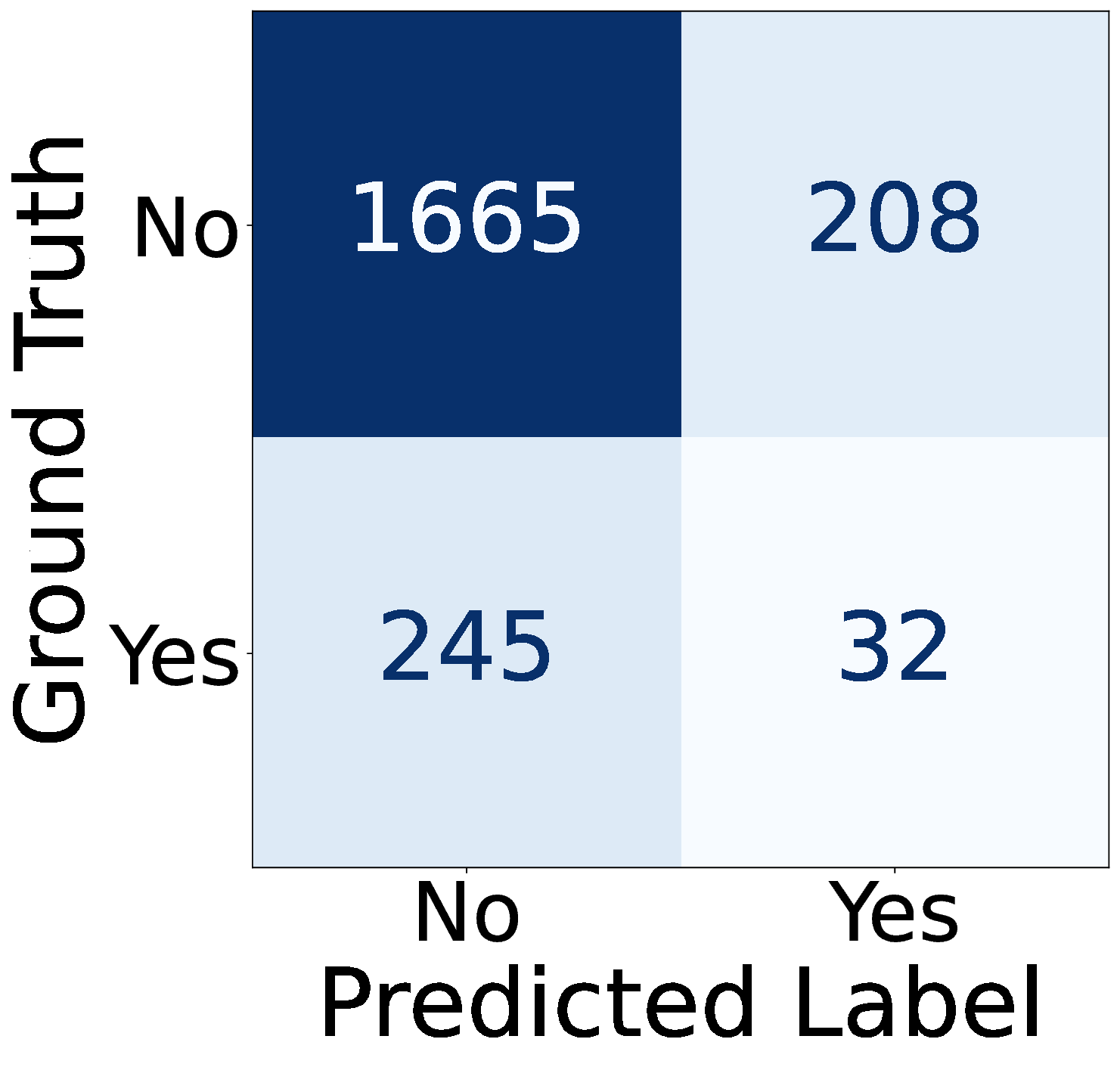, width=34mm}
        \subcaption{implicit}
        \label{subfigure:gpt-3.5_implicit_confusion_matrix}
      \end{minipage} &
      \begin{minipage}[t]{0.45\hsize}
        \centering
        \epsfig{file=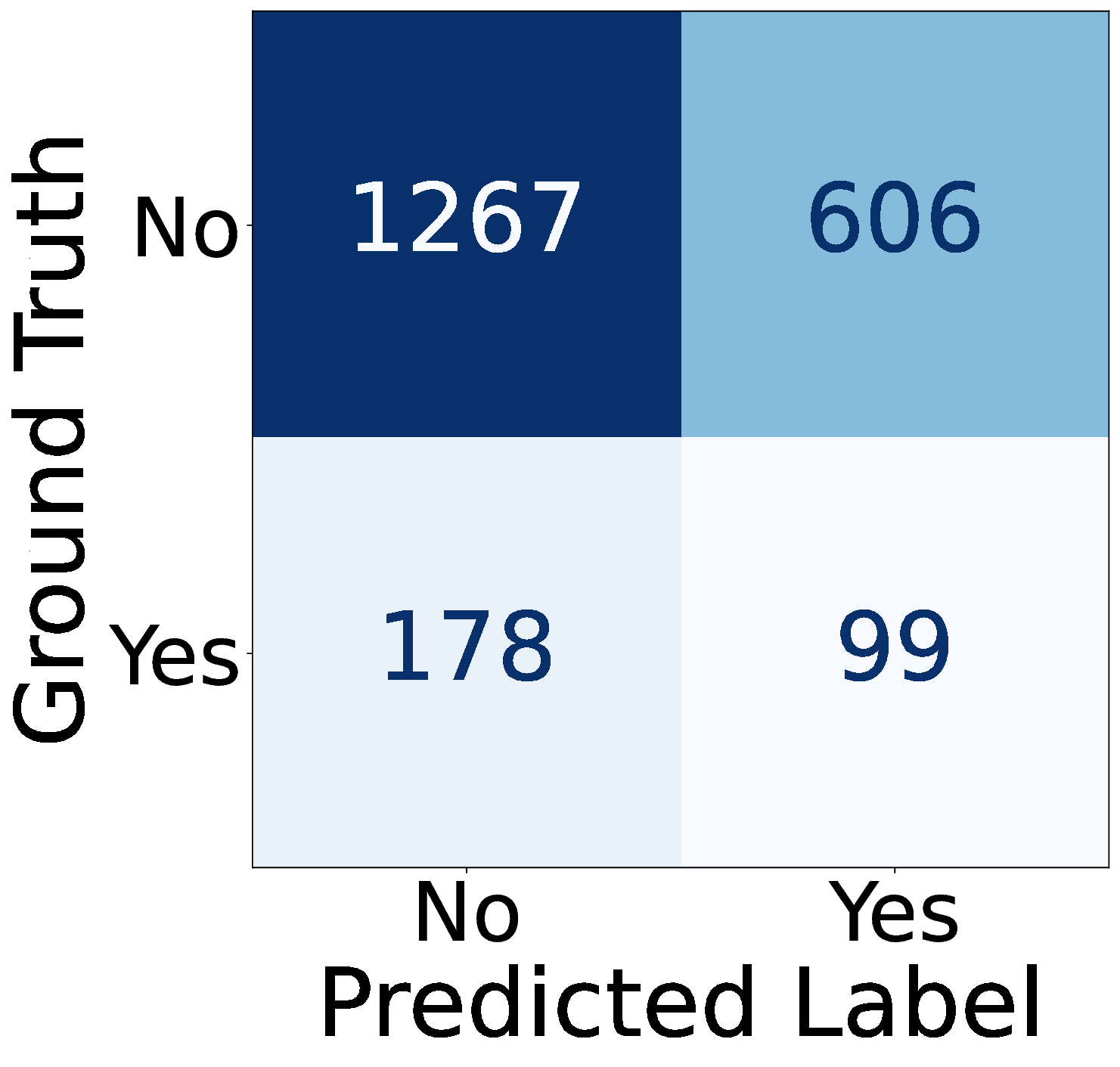, width=34mm}
        \subcaption{explicit}
        \label{subfigure:gpt-3.5_explicit_confusion_matrix}
      \end{minipage} \\

      \begin{minipage}[t]{0.45\hsize}
        \centering
        \epsfig{file=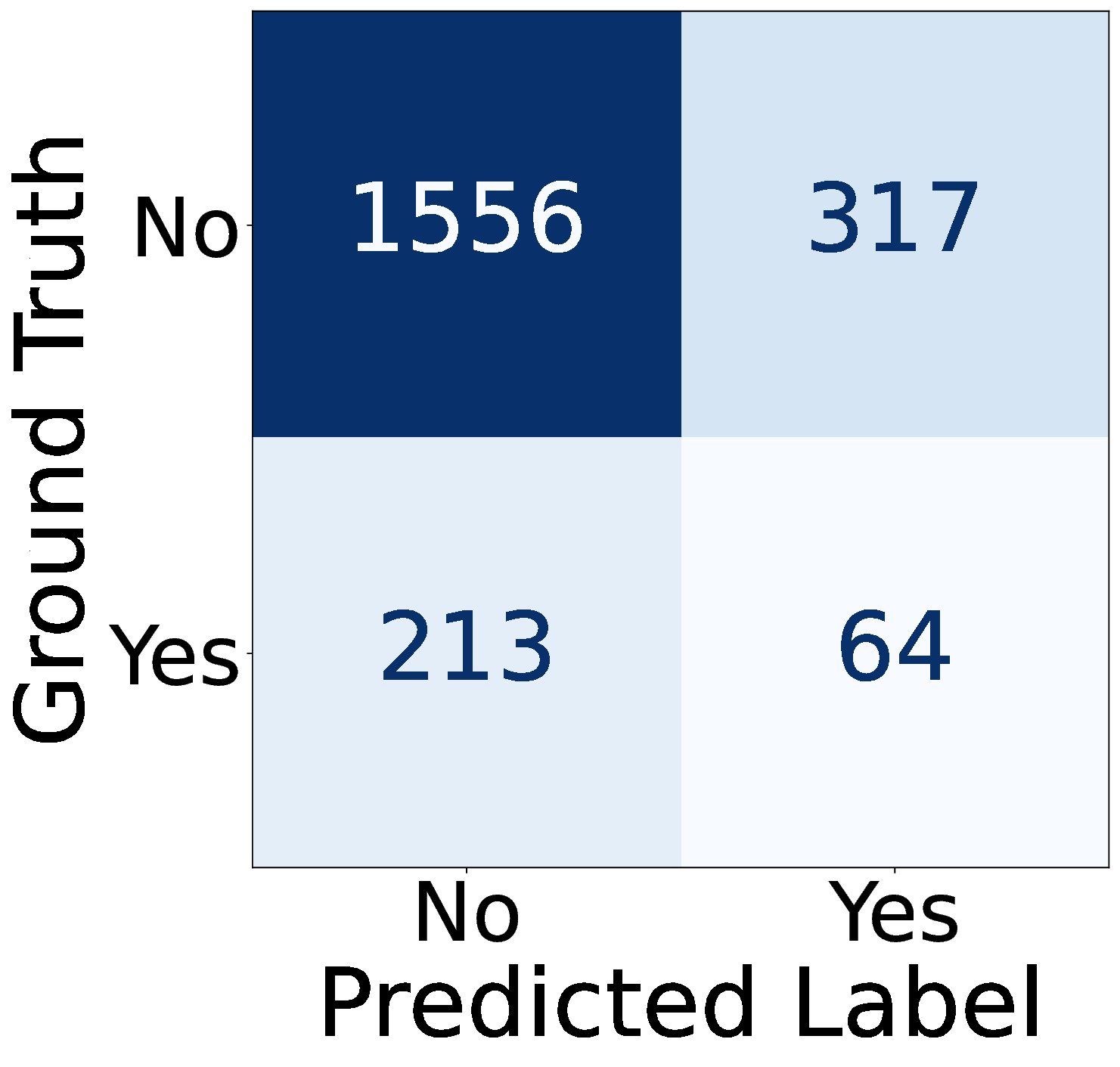, width=34mm}
        \subcaption{implicit w/ genres}
        \label{subfigure:gpt-3.5_implicit_with_genres_confusion_matrix}
      \end{minipage} &
      \begin{minipage}[t]{0.45\hsize}
        \centering
        \epsfig{file=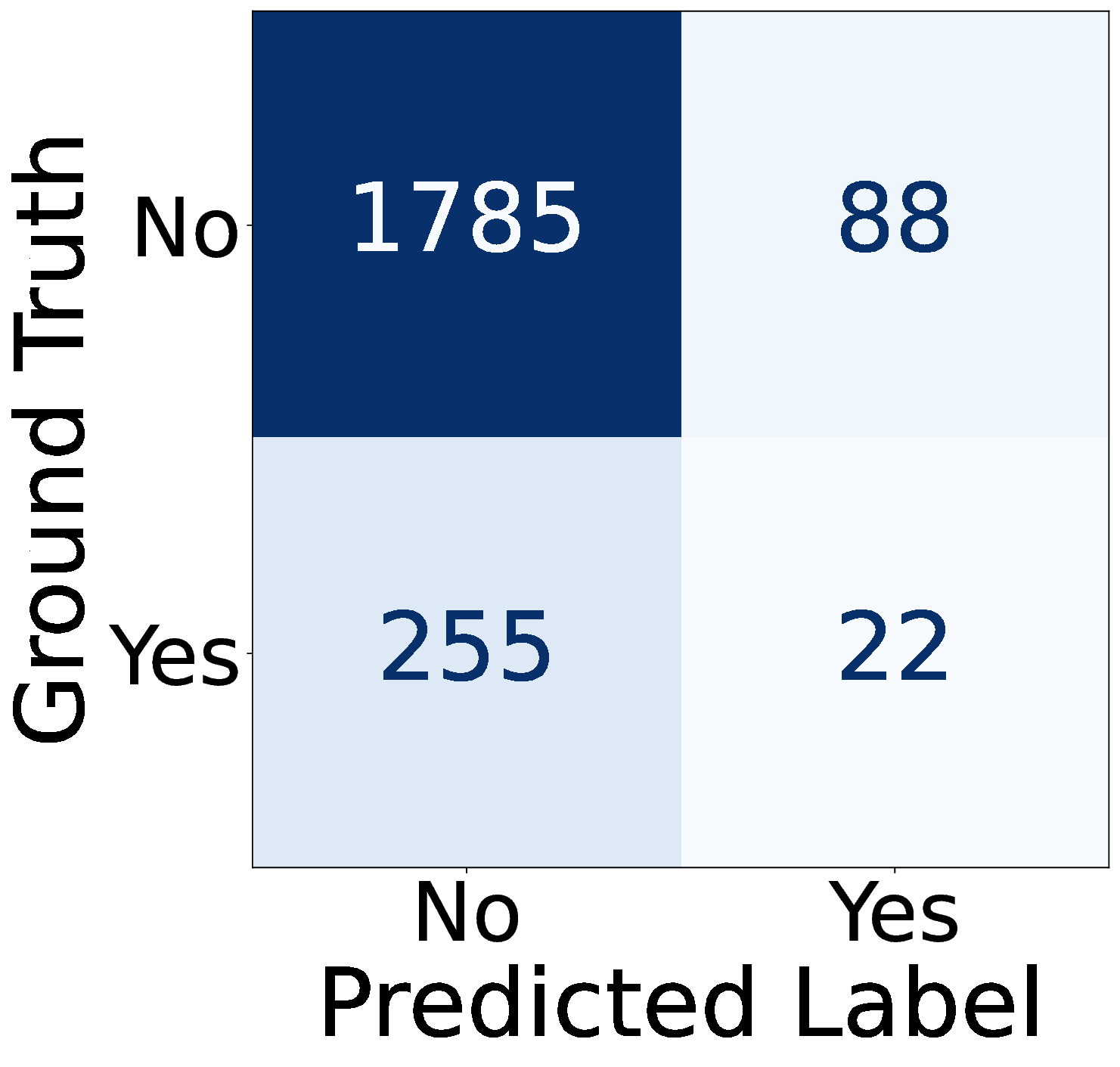, width=34mm}
        \subcaption{explicit w/ genres}
        \label{subfigure:gpt-3.5_explicit_with_genres_confusion_matrix}
      \end{minipage}
    \end{tabular}
    \caption{Confusion Matrix - GPT-3.5}
    \label{figure:gpt-3.5_confusion_matrix}
\end{figure}

%% file: table/accuracy_length_history.tex
\begin{table}[t]
  \centering
  \caption{Classification performance comparison for different lengths of rating history - GPT-3.5 (implicit w/ genres)}
  \label{table:accuracy_length_history}
  \begin{tabular}{cccccc}
    \hline
    $n$ & Accuracy & Precision & Recall & F1-score \\
    \hline
    \hline
    3 & 0.628 & 0.506 & 0.513 & 0.478 \\
    5 & 0.647 & 0.516 & 0.532 & 0.494 \\
    10 & \textbf{0.753} & \textbf{0.524} & 0.531 & \textbf{0.525} \\
    20 & 0.611 & 0.521 & \textbf{0.543} & 0.485 \\
    50 & 0.595 & 0.516 & 0.534 & 0.475 \\
    \hline
  \end{tabular}
\end{table}

%% file: table/pr-auc_logistic_regression.tex
\begin{table}[t]
  \centering
  \caption{PR-AUC with respect to the output of each LLM-based method and logistics regression model}
  \label{table:pr-auc}
  \begin{tabular}{lc}
    \hline
    Method & PR-AUC \\
    \hline\hline
    Llama2-13B-Chat (implicit) & 0.574 \\
    Llama2-13B-Chat (explicit) & 0.925 \\
    Llama2-13B-Chat (implicit w/ genres) & 0.138 \\
    Llama2-13B-Chat (explicit w/ genres) & 0.981 \\
    \hline
    GPT-3.5 (implicit) & 0.274 \\
    GPT-3.5 (explicit) & 0.551 \\
    GPT-3.5 (implicit w/ genres) & 0.220 \\
    GPT-3.5 (explicit w/ genres) & 0.072 \\
    \hline
    GPT-4 (implicit) & 0.955 \\
    GPT-4 (explicit) & 0.884 \\
    GPT-4 (implicit w/ genres) & 0.629 \\
    GPT-4 (explicit w/ genres) & 0.577 \\
    \hline
  \end{tabular}
\end{table}

%% file: table/logistic_regression_result.tex
\begin{table*}[t]
  \centering
  \caption{Result of logistic regression (Limit to methods with PR-AUC $>$ 0.9)}
  \label{table:log_reg}
  \begin{tabular}{lccccc}
    \hline
    Method & & const & $\hat{r}$ & $prof$ & $unpop$ \\
    \hline\hline
    Llama2-13B-Chat (explicit) & Coef & 2.318 & -0.039 & 0.247 & 0.097 \\
    & $p$-value & \underline{0.000} & 0.614 & \underline{0.001} & 0.324 \\
    & 95\%CI & $[2.167, 2.468]$ & $[-0.190, 0.112]$ & $[0.100, 0.394]$ & $[-0.095, 0.289]$\\
    \hline
    Llama2-13B-Chat & Coef & 3.576 & -0.022 & 0.603 & -0.049 \\
    (explicit w/ genres) & $p$-value & \underline{0.000} & 0.850 & \underline{0.000} & 0.666 \\
    & 95\%CI & $[3.300, 3.852]$ & $[-0.252, 0.207]$ & $[0.365, 0.841]$ & $[-0.271, 0.173]$ \\
    \hline
    GPT-4 (implicit) & Coef & 2.702 & 0.206 & 0.190 & -0.401 \\
    & $p$-value & \underline{0.000} & \underline{0.009} & \underline{0.028} & \underline{0.000} \\
    & 95\%CI & $[2.524, 2.880]$ & $[0.052, 0.360]$ & $[0.021, 0.360]$ & $[-0.515, -0.287]$ \\
    \hline
  \end{tabular}
\end{table*}

%% file: tex/section6.tex
\section{Conclusion} \label{sec:conclusion}
In this study, we tackled the issue of whether users perceive recommended items as serendipitous in RSs using LLMs and proposed a prompt for this purpose.
This prompt can be given to LLMs along with auxiliary information such as the user ratings and the genre of the item.
The experimental findings are as follows:
(1) LLM-based serendipitous assessment performed equivalently or better than the baseline method (RQ2).
(2) However, the agreement with human assessments was not as high (RQ1), suggesting the need for further improvement before LLMs can replace human assessments.
Moreover, we observed performance differences based on the LLMs used and that the classification performance of assessment is improved by providing LLMs with the genre information of the items.
We also discovered that the performance is negatively impacted when the prompt includes too few or too many user-rated items and that LLMs' outputs with good classification performance cannot be interpreted using baseline methods.

In future studies, we aim to enhance the prompts to improve performance and output the inference process.
Furthermore, we aim to utilize the proposed method for data augmentation on serendipity and for serendipitous assessments of recommendation lists for multiple RSs.
We believe that data augmentation is valuable and must be addressed.
In addition, we aim to model the user's serendipity as a recommender agent, and not solely in the assessment of recommendation results, as an application of LLMs.

%% file: tex/acknowledgement.tex
\acknowledgements
This work was supported by JSPS KAKENHI Grant Numbers JP21H03553, JP22H03698.

%% file: tex/appendix.tex
\appendix
\section{Generation of Serendipitous Ground Truth Labels}
For each of the 2,150 responses in the Serendipity-2018 dataset, we assigned a binary label indicating whether the user perceived the item as serendipitous.
Specifically, based on the responses to each question in \textbf{Table \ref{table:questions}}, the label is defined as positive if one of the following six conditions is met and negative otherwise.

\begin{itemize}
    \item (Q1 $>$ 3) $\land$ (Q4 $>$ 3)
    \item (Q1 $>$ 3) $\land$ (Q5 $>$ 3)
    \item (Q1 $>$ 3) $\land$ (Q6 $>$ 3)
    \item (Q2 $>$ 3) $\land$ (Q4 $>$ 3)
    \item (Q2 $>$ 3) $\land$ (Q5 $>$ 3)
    \item (Q2 $>$ 3) $\land$ (Q6 $>$ 3)
\end{itemize}

%% file: tex/profiles.tex
\begin{profile}
  \Name{Yu Tokutake}
  \Affiliation{Department of Informatics, Graduate School of Informatics and Engineering, The University of Electro-Communications}
  \Address{1-5-1 Chofugaoka, Chofu, Tokyo 182-8585, Japan}
  \History{
  2019-2023 B.E., The University of Electro-Communications, Japan \\
  2023- M.E., The University of Electro-Communications, Japan
  }
  \Works{$\bullet$ Y. Tokutake and K. Okamoto, ``Serendipity-oriented Recommender System with Dynamic Unexpectedness Prediction,'' Proc. of 2023 IEEE Int. Conf. on Syst., Man, and Cybern., pp. 1247--1252, 2023.}
  % \Membership{$\bullet$ Your Learned Societies}
 \Photo[\epsfig{file=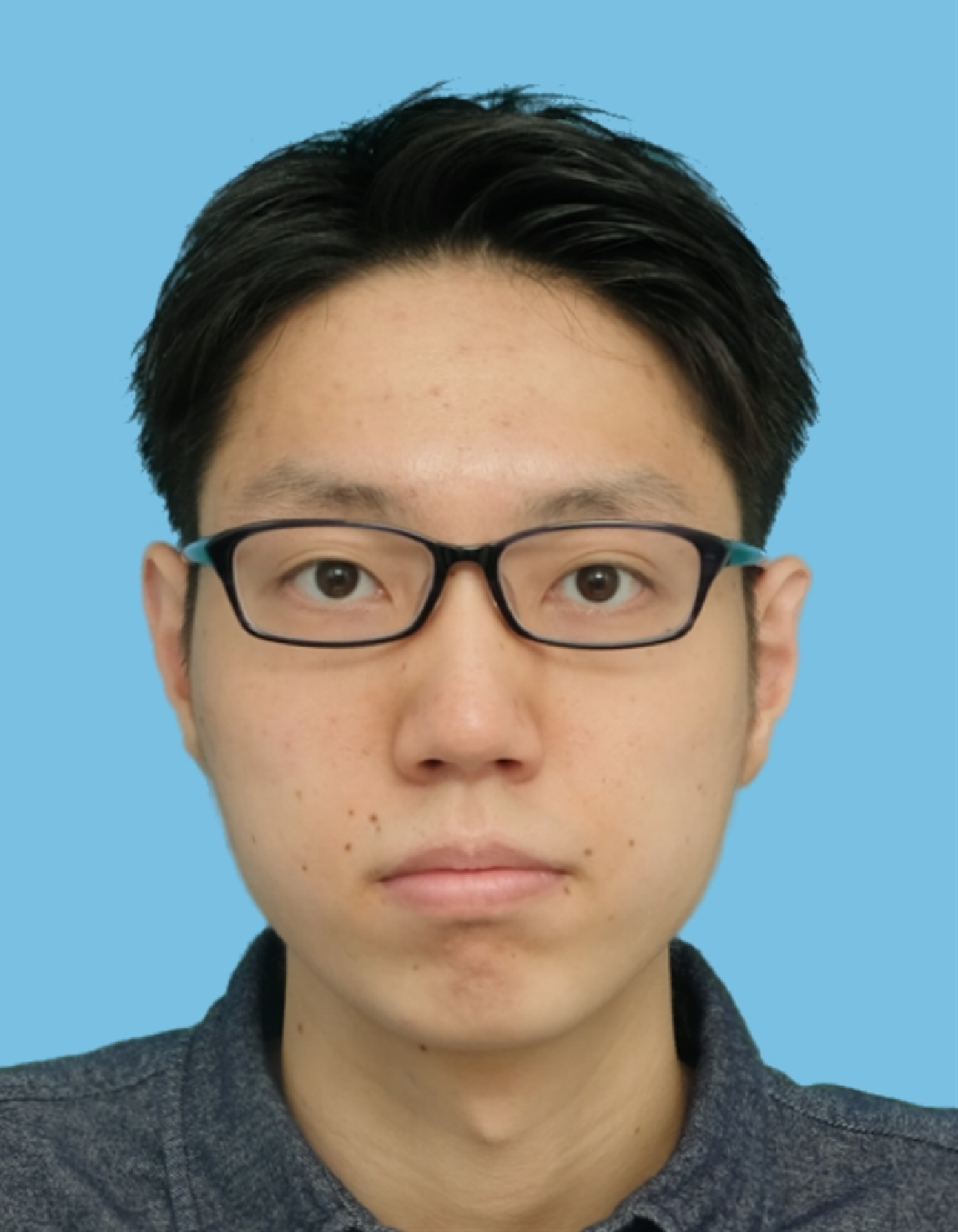,height=93.54pt, width=70.86pt}]
\end{profile}
\begin{profile}
  \Name{Kazushi Okamoto}
  \Affiliation{Department of Informatics, Graduate School of Informatics and Engineering, The University of Electro-Communications}
  \Address{1-5-1 Chofugaoka, Chofu, Tokyo 182-8585, Japan}
  \History{
    2002-2006 B.E., Kochi University of Technology, Japan \\
    2006-2008 M.E., Kochi University of Technology, Japan \\
    2008-2011 Dr. Eng., Tokyo Institute of Technology, Japan \\
    2011-2015 Assistant Professor, Chiba University, Japan \\
    2015-2020 Assistant Professor, The University of Electro-Communications, Japan \\
    2020- Associate Professor, The University of Electro-Communications, Japan
  }
  \Works{
    $\bullet$ K. Okamoto, ``Analysis of Influence Factors for Learning Outcomes with Bayesian Network,'' J. of Adv. Comput. Intell. Intell. Inform. Vol.22, No.6, 943--955, 2018. \\
    $\bullet$ K. Sugahara, K. Okamoto, ``Hierarchical Co-clustering with Augmented Matrices from External Domains,'' Pattern Recognit., Vol.142, pp.109657, 2023. \\
    $\bullet$ K. Sugahara, Kazushi Okamoto: ``Hierarchical Matrix Factorization for Interpretable Collaborative Filtering,'' Pattern Recognit. Lett., Vol.180, pp.99--106, 2024.
  }
  \Membership{
    $\bullet$ Japan Society for Fuzzy Theory and Systems (SOFT) \\
    $\bullet$ The Institute of Electronics, Information and Communication Engineers (IEICE) \\
    $\bullet$ Information Processing Society of Japan (IPSJ) \\
    $\bullet$ The Japanese Society for Artificial Intelligence (JSAI)
  }
  \Photo[\epsfig{file=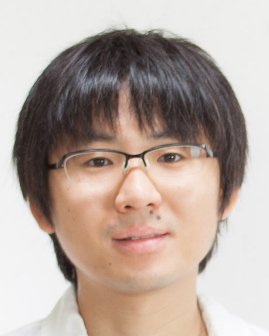,height=93.54pt, width=70.86pt}]
\end{profile}